\documentclass[aps,pra,floatfix,showpacs,preprint,onecolumn]{revtex4-1}

\usepackage{graphicx}
\usepackage{amsmath, amsfonts, amssymb, bm}

\usepackage{amstext}
\usepackage{mathrsfs}
\usepackage{color}

\begin{document}
\title{First-Order Strong-Field QED Processes in a Tightly Focused Laser Beam}
\author{A.~Di Piazza}
\email{dipiazza@mpi-hd.mpg.de}
\affiliation{Max-Planck-Institut f\"ur Kernphysik, Saupfercheckweg 1, D-69117 Heidelberg, Germany}

\begin{abstract}
In [Phys. Rev. Lett. \textbf{117}, 213201 (2016)] we have determined the angular resolved and the total energy spectrum of a positron produced via nonlinear Breit-Wheeler pair production by a high-energy photon counterpropagating with respect to a tightly focused laser beam. Here, we first generalize the results in [Phys. Rev. Lett. \textbf{117}, 213201 (2016)] by including the possibility that the incoming photon is not exactly counterpropagating with respect to the laser field. As main focus of the present paper, we determine the photon angular resolved and total energy spectrum for the related process of nonlinear Compton scattering by an electron impinging into a tightly-focused laser beam. Analytical integral expressions are obtained under the realistic assumption that the energy of the incoming electron is the largest dynamical energy of the problem and that the electron is initially almost counterpropagating with respect to the laser field. The crossing symmetry relation between the two processes in a tightly focused laser beam is also elucidated.
\end{abstract}

\maketitle

\section{Introduction}

The emission of radiation by accelerated electric charges is one
of the most fundamental processes in physics. While classically
the emission of radiation is a continuous process \cite{Jackson_b_1975},
quantum mechanically it has a discrete nature meaning that
the radiation is emitted as quanta, called photons \cite{Landau_b_4_1982}.
The process of photon emission by a massive charged particle, an electron for definiteness,
cannot occur in vacuum due to energy-momentum conservation. However,
if the electron interacts with a background electromagnetic field,
the latter can provide the missing energy and momentum and the emission
of photons can occur. If the background electromagnetic field is sufficiently strong
that during the emission process: a) it is not altered by the emission 
process itself and b) many of its photons interact with the electron,
then the so-called Furry picture can be efficiently employed to calculate the
emission process probability for a relatively large class of background fields 
by taking into account exactly the background field itself in the calculations \cite{Furry_1951,Landau_b_4_1982,Fradkin_b_1991,Dittrich_b_1985}.
In the Furry picture the electron states and propagator are determined 
exactly in the external electromagnetic field, meaning
that they are obtained by solving the Dirac equation and the corresponding equation
for the propagator including the background field. After that the obtained
``dressed'' states and ``dressed'' propagator can be employed within the conventional Feynman
approach to determine the probabilities of QED processes
by accounting perturbatively for the interaction between the electron-positron
field and the photon field (apart from this, depending on the structure of the background field,
the theory of QED in the presence of a strong background field may have 
qualitatively additional different features like, for example, the instability 
of the vacuum under
electron-positron pair production \cite{Fradkin_b_1991}). 

Below we are interested in the case where the background electromagnetic field
is a laser field, typically with an optical frequency, corresponding to a wavelength
of the order of one micrometer. In fact, present high-power 
optical laser facilities have reached very high intensities of the order 
of $I_0\sim 10^{22}\;\text{W/cm$^2$}$ \cite{Yanovsky_2008} and upcoming 
10-PW facilities aim at $I_0\sim 10^{23}\;\text{W/cm$^2$}$ \cite{APOLLON_10P}. 
The requirement of high intensity is related to the importance of so-called nonlinear
quantum effects like, for example, the recoil undergone by the electron in the
emission of photons. This is a pure quantum effect and, as other QED effects in a strong
laser field (approximated here as a plane wave), it is controlled by the so-called 
quantum nonlinearity parameter $\chi_0=|e|\sqrt{|(p_{\mu}F_0^{\mu\nu})^2|}/m^3$ \cite{Mitter_1975,Ritus_1985,Ehlotzky_2009,Reiss_2009,Di_Piazza_2012,Dunne_2014}. Here, $e<0$ and $m$ are the electron charge and mass, respectively, $p^{\mu}=(\varepsilon,\bm{p})$ is the initial four-momentum of the electron, and $F_0^{\mu\nu}=(\bm{E}_0,\bm{B}_0)$ is a measure of the amplitude of the laser field, with $E_0=B_0=F_0$ (units with $4\pi\epsilon_0=\hbar=c=1$ are employed throughout). If one denotes with $n^{\mu}=(1,\bm{n})$ ($n^2=0$) the four-momentum of a laser photon in units of its energy and with $A_0^{\mu}=(0,\bm{A}_0)=(0,-\bm{E}_0/\omega_0)$ the amplitude of the four-vector potential, where $\omega_0$ is the central angular frequency of the laser, the parameter $\chi_0$ can be written as $\chi_0=((np)/m)F_0/F_{cr}$, where $F_{cr}=m^2/|e|=1.3\times 10^{16}\;\text{V/cm}=4.4\times 10^{13}\;\text{G}$ is the so-called critical field of QED \cite{Mitter_1975,Ritus_1985,Ehlotzky_2009,Reiss_2009,Di_Piazza_2012,Dunne_2014}. The above expression of the parameter $\chi_0$ sets the field scale $F_{cr}$ as the typical scale where non-linear QED effects, like the importance of the recoil undergone by the electron when it emits a photon, are significant. The same expression, however, also indicates that even for laser field amplitudes much below the critical value, nonlinear QED effects can be important ($\chi_0\gtrsim 1$) if, for example, one employs ultrarelativistic electrons which initially counterpropagate with respect to the laser field. In fact, from a physical point of view the relevant quantity here is the external field amplitude in the initial rest frame of the electron $((np)/m)F_0$ and for an ultrarelativistic electron initially counterpropagating with the laser field it is $(np)/m\approx 2\varepsilon/m\gg 1$. Since the critical amplitude $F_{cr}$ corresponds to a laser intensity $I_{cr}=4.6\times 10^{29}\;\text{W/cm$^2$}$, it is clear that the strong-field QED regime, where nonlinear QED effects become essential, can be entered nowadays and in the near future only by employing ultrarelativistic electron beams. It is worth pointing out that conventional accelerators have provided electron beams with energies of the order of $50\;\text{GeV}$ \cite{Bula_1996,Burke_1997} whereas modern accelerators based on the laser-wakefield acceleration technique have already reached energies of the order of $1\text{-}5\;\text{GeV}$ \cite{Leemans_2014}. Thus, having in mind the mentioned experimentally achieved laser intensities we can conclude that present technology allows in principle for entering the strong-field QED regime. For the sake of completeness, we have to remind that another requirement for entering the strong-field regime is the importance of nonlinear effects in the amplitude of the external field. In fact, if the field is so weak that during the process under consideration an electron effectively interacts only with a single external-field photon then even the use of the Furry picture is redundant and essentially the amplitudes of vacuum QED can be employed to calculate transition probabilities and rates. In the case of a laser field the importance of nonlinear effects with respect to the laser amplitude is related to the energy that the laser field can transfer to an electron in the typical QED length $\lambda_C=1/m=3.9\times 10^{-11}\;\text{cm}$ (Compton wavelength) and it is controlled by the parameter $\xi_0=|e|E_0/m\omega_0$ \cite{Mitter_1975,Ritus_1985,Ehlotzky_2009,Reiss_2009,Di_Piazza_2012,Dunne_2014}. If this parameter is larger than unity, in fact, the laser field can transfer an energy corresponding to many laser photons to an electron on a Compton wavelength. Since the threshold $\xi_0=1$ corresponds to an optical ($\omega_0\sim 1\;\text{eV}$) laser intensity of the order of $10^{18}\;\text{W/cm$^2$}$, it is customary to consider the highly nonlinear regime where $\xi_0\gg 1$ (see \cite{Dinu_2016} for a recent study where also interesting features in the regime $\xi_0\sim 1$ are investigated).

The process of the emission of a single photon by an electron in the field of a plane wave (nonlinear single Compton scattering) has been studied since the sixties \cite{Goldman_1964,Nikishov_1964} and, due to the fast development of laser technology, has received again a large attention in the last years \cite{Ivanov_2004,Boca_2009,Harvey_2009,Mackenroth_2010,Boca_2011,Mackenroth_2011,Seipt_2011,Seipt_2011b,Krajewska_2012,Seipt_2013,Krajewska_2014,Wistisen_2014,Harvey_2015,Seipt_2016,Seipt_2016b,Angioi_2016,Harvey_2016b}, with emphasis on effects related to the laser duration, laser polarization, spin of the participating electron and so on (see also the reviews \cite{Mitter_1975,Ritus_1985,Ehlotzky_2009,Reiss_2009,Di_Piazza_2012}). Moreover, the process of double photon emission in the field of a plane wave (nonlinear double Compton scattering) has also been investigated recently \cite{Seipt_2012,Mackenroth_2013,King_2015}. 

In all above mentioned studies in the full quantum regime the laser field has been approximated as a plane-wave, which allows one to solve analytically the Dirac equation and obtain the exact electron states and propagators (Volkov states and propagator, respectively) \cite{Volkov_1935,Landau_b_4_1982} and to use them within the Furry picture. However, present and upcoming laser facilities may realistically reach the high intensities required to enter the strong-field QED regime only by tightly focusing the laser energy both in space and time. Although arbitrary laser pulse shapes in time can be accounted for within the plane-wave approximation of the laser field, space focusing goes beyond this approximation. Recently, effects of the laser spatial focusing in Compton and Thomson scattering (the latter process corresponds to the classical emission of radiation, where quantum effects like recoil can be neglected) have been recently investigated numerically in \cite{Li_2015} and in \cite{Harvey_2016}, respectively. Also, analytical expressions of scalar wave functions based on the Wentzel-Kramers-Brillouin (WKB) approximation have been determined in \cite{Heinzl_2016} for a specific class of background fields depending on the space-time coordinates via the quantity $(nx)$ like a plane wave but generalizing from lightlike $n^{\mu}$ to arbitrary $n^{\mu}$. Moreover, the dynamics of a scalar particle in a background formed by two counter-propagating plane waves both in the classical and in the quantum regime has been recently studied in \cite{King_2016}. In \cite{Di_Piazza_2014,Di_Piazza_2015} we have started to investigate a regime of laser-electron interaction which is relevant for presently and forthcoming experiments in strong-field QED and where it was possible to determine analytically the wave-functions \cite{Di_Piazza_2014,Di_Piazza_2015} and the propagator \cite{Di_Piazza_2015} for an electron in the presence of a background electromagnetic field of virtually arbitrary space-time structure having in mind the case of a tightly-focused laser beam. In this regime the involved charged particles (electrons and positrons) are assumed to be (almost) counterpropagating with respect to the laser field (the meaning of ``almost'' is clarified in the next section). Although other geometries can be easily implemented (see \cite{Di_Piazza_2014}), the counterpropagating setup, as we have indicated above, is the one featuring the largest value of the quantum nonlinearity parameter $\chi_0$ at a given field amplitude and incoming electron energy. Moreover, the longitudinal momentum of the involved charged particles is much larger than the typical transverse momentum scale $m\xi_0$ in the laser field such that the charged particles are barely deflected by the laser field and their energy scale is determined by the longitudinal momentum. This has allowed us to solve the Dirac equation within the WKB approximation but by keeping next-to-leading order terms, which are essential already to reproduce the results in the plane wave limit. It is important to stress that the approximation $\varepsilon\gg m\xi_0$ is automatically satisfied for present and upcoming experimental conditions. In fact, let us assume for the sake of definiteness that an ultrarelativistic electron initially counterpropagates with respect to the laser beam. In order to enter the strong-field QED regime (say at $\chi_0>1$), by assuming the laser to be a Ti:Sapphire laser ($\omega_0=1.55\;\text{eV}$) and to have a soon feasible intensity of $I_0\sim 10^{23}\;\text{W/cm$^2$}$  \cite{APOLLON_10P} (corresponding to $\xi_0=150$), it is necessary that $\varepsilon\gtrsim 500\;\text{MeV}$ such that it is $\varepsilon/m\approx 10^3$. In \cite{Di_Piazza_2016} we have shown that the wave functions found in \cite{Di_Piazza_2014,Di_Piazza_2015} can indeed be employed to obtain relatively compact integral analytical expressions for the angular resolved and the total energy spectrum of positrons produced in the head-on collision of a high-energy photon with a strong and tightly focused laser beam. The spectra in \cite{Di_Piazza_2016} are conveniently expressed as functions of the external background field and have been shown to be in agreement with the corresponding results obtained by means of the operator technique in the quasiclassical approximation \cite{Baier_b_1998} (see also \cite{Akhiezer_b_1996}), although, in general, in the latter approach the angular resolved and the total energy spectra are expressed in terms of the electron trajectory, which has to be determined separately. Here, we first generalize the findings in \cite{Di_Piazza_2016} to include the possibility that the incoming photon is not exactly counterpropagating with respect to the 
laser beam. Then, we focus on the analogous study of nonlinear single Compton scattering in a tightly focused laser beam. In addition, we will elucidate how these two first-order strong-field QED processes are related by the crossing symmetry \cite{Landau_b_4_1982}.

\section{Nonlinear Breit-Wheeler pair production}

In this section we generalize the results obtained in \cite{Di_Piazza_2016} including the possibility that the incoming photon is not exactly counterpropagating with respect to the laser beam. As in \cite{Di_Piazza_2016}, we assume that the laser field is described by the four-vector potential $A^{\mu}(x)$ in the Lorentz gauge $\partial_{\mu}A^{\mu}(x)=0$. For the sake of definiteness, we consider a laser beam whose focal plane corresponds to the $x\text{-}y$ plane and whose wave vector at the center of the focal area points along the negative $z$ direction (below we summarize these properties by saying that the laser beam propagates along the negative $z$ direction). A concrete realistic form of the background electromagnetic field which can be studied with the present formalism and its main features are presented in the Appendix A. The chosen setup described above suggests to introduce the light-cone coordinates $T=(t+z)/2$, $\bm{x}_{\perp}=(x,y)$, and $\phi=t-z$ for a generic four-position $x^{\mu}=(t,x,y,z)$. Analogously, it is convenient to introduce the light-cone components $v_{\pm}=(v^0\pm v_z)/2^{(1\pm 1)/2}$ and $\bm{v}_{\perp}=(v_x,v_y)$ for an arbitrary four-vector $v^{\mu}=(v^0,v_x,v_y,v_z)$. The four-momentum and the polarization four-vector of the incoming photon are indicated as $k^{\mu}=(\omega,\bm{k})$ ($k^2=0$) and $e^{\mu}_{k,l}$ ($l=1,2$), respectively (the four-vectors $e^{\mu}_{k,l}$ are considered real implying that the incoming photon is assumed to be linearly polarized). The incoming photon is ``almost'' propagating along the positive $z$ direction meaning that $|\bm{k}_{\perp}|\lesssim m\xi_0\ll k_z\approx\omega$ (see in particular \cite{Di_Piazza_2015}). Concerning the final electron-positron pair, it is convenient here to indicate as $p^{\mu}=(\varepsilon,\bm{p})$ ($p^2=m^2$) and $s=1,2$ the four-momentum and the spin quantum number of the positron and as $p^{\prime\mu}=(\varepsilon',\bm{p}')$ ($p^{\prime2}=m^2$) and $s'=1,2$ the corresponding quantities for the electron. Although this is the opposite notation as that we have employed in \cite{Di_Piazza_2016}, it will simplify the comparison with the results in nonlinear single Compton scattering and the discussion on the crossing symmetry. Thus, for the sake of clarity, we will report here also some formulas, which differ from the corresponding ones in \cite{Di_Piazza_2016} by the exchange of $p^{\mu}$ with $p^{\prime\mu}$ and of $s$ with $s'$. The amplitude of the nonlinear Breit-Wheeler pair production at the leading order within the Furry picture is given by
\begin{equation}
\label{S_BW}
S_{BW,fi}=-ie\sqrt{4\pi}\int d^4x\,\bar{\psi}^{(\text{out})}_{p',s'}(x)\frac{\hat{e}_{k,l}}{\sqrt{2\omega}}e^{-i(kx)}\psi^{(\text{out})}_{-p,-s}(x),
\end{equation}
where the hat indicates the contraction of a four-vector with the Dirac matrices $\gamma^{\mu}$, and where $\bar{\psi}=\psi^{\dag}\gamma^0$ for an arbitrary bispinor $\psi$ (a unit quantization volume is assumed). The out-states employed to evaluate the transition amplitude in Eq. (\ref{S_BW}) are those given in Eqs. (2)-(3) in \cite{Di_Piazza_2016}, i.e.:
\begin{equation}
\label{out_states}
\psi^{(\text{out})}_{\pm p,\pm s}(x)=e^{iS^{(\text{out})}_{\pm p}(x)}\bigg[1\pm\frac{e}{4 p_+}\hat{n}\hat{\mathcal{A}}^{(\text{out})}(\bm{x})\bigg] \frac{u_{\pm p,\pm s}}{\sqrt{2\varepsilon}},
\end{equation}
where
\begin{equation}
\begin{split}
S^{(\text{out})}_{\pm p}(x)=&\mp(p_+\phi+p_-T-\bm{p}_{\perp}\cdot\bm{x}_{\perp})+e\int_T^{\infty}d\tilde{T} A_-(\tilde{\bm{x}})\\
&+\frac{1}{p_+}\int_T^{\infty}d\tilde{T} \left[e(p\mathcal{A}^{(\text{out})}(\tilde{\bm{x}}))\mp
\frac{1}{2}e^2\mathcal{A}^{(\text{out})\,2}(\tilde{\bm{x}})\right],
\end{split}
\end{equation}
where $\mathcal{A}^{(\text{out}),\mu}(\bm{x})=(0,\bm{\mathcal{A}}^{(\text{out})}_{\perp}(\bm{x}),0)$, with
\begin{equation}
\label{A_out}
\bm{\mathcal{A}}^{(\text{out})}_{\perp}(\bm{x})=\bm{A}_{\perp}(\bm{x})-\bm{\nabla}_{\perp}\int_T^{\infty}d\tilde{T}A_-(\tilde{\bm{x}})=\int_T^{\infty}d\tilde{T}[\bm{E}_{\perp}(\tilde{\bm{x}})+\bm{z}\times\bm{B}_{\perp}(\tilde{\bm{x}})],
\end{equation}
and where $u_{\pm p,\pm s}$ are the positive-/negative-energy constant free bispinors \cite{Landau_b_4_1982} (the symbol $\bm{x}$ denotes the three coordinates $(T,\bm{x}_{\perp})$ and, correspondingly, $\tilde{\bm{x}}=(\tilde{T},\bm{x}_{\perp})$). We recall that under our approximations one can neglect the dependence of the background field on the variable $\phi$ and, for an appropriate choice of the initial conditions, evaluate the background field itself at $\phi\approx 0$ \cite{Di_Piazza_2014}. The reason is that, as it can be also ascertained from the classical motion of an ultrarelativistic charged particle along the positive $z$ direction, the quantity $\phi=t-z$ effectively scales as the square of the inverse of the energy of the particle in the relevant integration region (see also \cite{Di_Piazza_2014}). Thus, having in mind, for example, the expression of the background electromagnetic field discussed in the Appendix A, in order to perform concrete calculations, one has to replace $t\approx z\approx T$ in the expression of the background field, whereas the transverse coordinates $x$ and $y$ correspond to the two-dimensional vector $\bm{x}_{\perp}$.

Since the only difference with respect to the results in \cite{Di_Piazza_2016}, apart from the mentioned notational one, is the four-momentum of the incoming photon, we can already write that Eqs. (7)-(8) in \cite{Di_Piazza_2016} become
\begin{equation}
\label{dN_2_BW}
\begin{split}
\frac{dN_{BW}}{d\varepsilon d\Omega_p}=&i\rho_{\Sigma,\gamma}\frac{\alpha\varepsilon}{16\pi^3\omega}\int \frac{d^3\bm{x}d^3\bm{x}'}{T_-}e^{i\Delta\Phi_{BW}(\bm{x},\bm{x}')}\left\langle m^2\left(\frac{\varepsilon'}{\varepsilon}+\frac{\varepsilon}{\varepsilon'}+4\right)+\frac{2i\varepsilon}{T_-}+\frac{\varepsilon'}{\varepsilon}\bigg\{\bm{p}_{\perp}\right.\\
&\left.-\frac{\varepsilon}{T_-}\Delta\bm{x}_{\perp,p}+e\frac{\omega}{\varepsilon'}\frac{1}{T_-}\left[\int_T^{\infty}d\tilde{T} \bm{\mathcal{A}}_{\perp}(\tilde{\bm{x}})-\int_{T'}^{\infty}d\tilde{T}' \bm{\mathcal{A}}_{\perp}(\tilde{\bm{x}}')\right]+e\frac{\omega}{\varepsilon'}\bm{\mathcal{A}}_{\perp,+}(\bm{x},\bm{x}')\right\}^2\\
&\left.-e^2\frac{(\varepsilon-\varepsilon')^2}{4\varepsilon\varepsilon'}\bm{\mathcal{A}}^2_{\perp,-}(\bm{x},\bm{x}')\right\rangle
\end{split}
\end{equation}
and
\begin{equation}
\label{Delta_Phi_BW}
\begin{split}
\Delta\Phi_{BW}(\bm{x},\bm{x}')=&\frac{\omega}{\varepsilon\varepsilon'}\frac{T_-}{2}m^2-\frac{T_-}{2\omega}\left(\bm{k}_{\perp}-\frac{\omega}{T_-}\bm{x}_{\perp,-}\right)^2+\frac{T_-}{2\varepsilon}\left(\bm{p}_{\perp}-\frac{\varepsilon}{T_-}\Delta\bm{x}_{\perp,p}\right)^2\\
&-\frac{\omega}{\varepsilon\varepsilon'}\frac{e^2}{2}\left\{\frac{1}{T_-}\left[\int_T^{\infty}d\tilde{T} \bm{\mathcal{A}}_{\perp}(\tilde{\bm{x}})-\int_{T'}^{\infty}d\tilde{T}' \bm{\mathcal{A}}_{\perp}(\tilde{\bm{x}}')\right]^2\right.\\
&+\int_T^{\infty}d\tilde{T} \bm{\mathcal{A}}^2_{\perp}(\tilde{\bm{x}})-\int_{T'}^{\infty}d\tilde{T}' \bm{\mathcal{A}}^2_{\perp}(\tilde{\bm{x}}')\Bigg\}.
\end{split}
\end{equation}
In these equations we have introduced the differential positron solid angle $d\Omega_p\approx d^2\bm{p}_{\perp}/\varepsilon^2$ (indicated as $d\Omega$ in \cite{Di_Piazza_2016}), the number of incoming photons per unit surface $\rho_{\Sigma,\gamma}$ (defined as $\rho_{\Sigma}$ in \cite{Di_Piazza_2016}), the final electron energy $\varepsilon'=\omega-\varepsilon$, and the quantities $T_{\pm}=(T\pm T')/2^{(1\pm 1)/2}$, $\bm{x}_{\perp,\pm}=(\bm{x}_{\perp}\pm \bm{x}'_{\perp})/2^{(1\pm 1)/2}$, $\bm{\mathcal{A}}_{\perp,\pm}(\bm{x},\bm{x}')=[\bm{\mathcal{A}}_{\perp}(\bm{x})\pm\bm{\mathcal{A}}_{\perp}(\bm{x}')]/2^{(1\pm 1)/2}$, $\Delta\bm{x}_{\perp,p}=\bm{x}_{\perp,-}+(e/\varepsilon)\big[\int_T^{\infty}d\tilde{T} \bm{\mathcal{A}}_{\perp}(\tilde{\bm{x}})-\int_{T'}^{\infty}d\tilde{T}' \bm{\mathcal{A}}_{\perp}(\tilde{\bm{x}}')\big]$, whereas the symbol $\tilde{\bm{x}}'$ denotes the three coordinates $(\tilde{T},\bm{x}_{\perp}')$.

The above Eq. (\ref{dN_2_BW}) can be easily integrated with respect to the final transverse positron momentum $d^2\bm{p}_{\perp}\approx \varepsilon^2d\Omega_p$ because the integral is Gaussian as the phase there contains at the highest quadratic terms in $\bm{p}_{\perp}$. Since the pre-exponent also contains $\bm{p}_{\perp}$ (up to the second power), we need the identities (see, e.g., \cite{Gradshteyn_b_2000})
\begin{align}
\label{G_0}
\mathcal{I}_0(a)&=\int \frac{d^2\bm{z}}{(2\pi)^2}e^{ia\bm{z}^2}=\int_0^{\infty} \frac{ds}{4\pi}e^{ias}=\frac{i}{4\pi a},\\
\label{G_2}
\mathcal{I}_2(a)&=\int \frac{d^2\bm{z}}{(2\pi)^2}\,\bm{z}^2e^{ia\bm{z}^2}=-i\frac{d\mathcal{I}_0(a)}{da}=-\frac{1}{4\pi a^2}
\end{align}
for any two-dimensional real vector $\bm{z}=(z_1,z_2)$ and for any constant $a$ with $\text{Im}(a)>0$. By employing these identities in Eq. (\ref{dN_2_BW}), the result for the total positron energy spectrum is
\begin{equation}
\label{dN_e_BW}
\begin{split}
\frac{dN_{BW}}{d\varepsilon}=&-\rho_{\Sigma,\gamma}\frac{\alpha}{8\pi^2\omega}\int \frac{d^3\bm{x}d^3\bm{x}'}{T^2_-}\exp\left\langle i\left\{\frac{m^2}{2}\frac{\omega}{\varepsilon\varepsilon'}T_--\frac{T_-}{2\omega}\left(\bm{k}_{\perp}-\frac{\omega}{T_-}\bm{x}_{\perp,-}\right)^2\right.\right.\\
&-\frac{\omega}{\varepsilon\varepsilon'}\frac{e^2}{2}\frac{1}{T_-}\left[\int_T^{\infty}d\tilde{T} \bm{\mathcal{A}}_{\perp}(\tilde{\bm{x}})-\int_{T'}^{\infty}d\tilde{T}' \bm{\mathcal{A}}_{\perp}(\tilde{\bm{x}}')\right]^2\\
&\left.\left.-\frac{\omega}{\varepsilon\varepsilon'}\frac{e^2}{2}\left[\int_T^{\infty}d\tilde{T} \bm{\mathcal{A}}^2_{\perp}(\tilde{\bm{x}})-\int_{T'}^{\infty}d\tilde{T}' \bm{\mathcal{A}}^2_{\perp}(\tilde{\bm{x}}')\right]\right\}\right\rangle\left\langle m^2\left(\frac{\varepsilon'}{\varepsilon}+\frac{\varepsilon}{\varepsilon'}+4\right)+\frac{2i\omega}{T_-}\right.\\
&\left.+\frac{\omega^2}{\varepsilon\varepsilon'}e^2\left\{\frac{1}{T_-}\left[\int_T^{\infty}d\tilde{T} \bm{\mathcal{A}}_{\perp}(\tilde{\bm{x}})-\int_{T'}^{\infty}d\tilde{T}' \bm{\mathcal{A}}_{\perp}(\tilde{\bm{x}}')\right]+\bm{\mathcal{A}}_{\perp,+}(\bm{x},\bm{x}')\right\}^2\right.\\
&\left.-e^2\frac{(\varepsilon-\varepsilon')^2}{4\varepsilon\varepsilon'}\bm{\mathcal{A}}^2_{\perp,-}(\bm{x},\bm{x}')\right\rangle.
\end{split}
\end{equation}
As we have already pointed out in \cite{Di_Piazza_2016}, it is convenient at this point to consider the integrals in the transverse coordinates and to pass to the variables $\bm{x}_{\perp,\pm}=(\bm{x}_{\perp}\pm \bm{x}'_{\perp})/2^{(1\pm 1)/2}$. In fact, one then realizes that the transverse formation length of the process is of the order of the Compton wavelength ($\lambda_C=3.9\times 10^{-11}\;\text{cm}\sim 10^{-7}\;\text{$\mu$m}\sim 10^{-4}\;\text{nm}$). Now, we have in mind applications where the tightly focused laser field is either an optical field, which varies in space on scales of the order of one micrometer or an x-ray field, which vary on scales of the order of one nanometer. Thus, one can expand the external field around $\bm{x}_{\perp,+}$ and neglect there the difference between $\bm{x}_{\perp}$ and $\bm{x}'_{\perp}$, as the corrections will be proportional to the small parameter $\lambda_C/\lambda_0\sim\omega_0/m$, with $\lambda_0=2\pi/\omega_0$ being the central laser wavelength (needless to say the approximation works even better for lower-frequency lasers like terahertz lasers). As we have noticed in \cite{Di_Piazza_2016}, this is also justified at the leading order in $1/\omega$ as $\lambda_C/\lambda_0\sim (\kappa_0/\xi_0)(m/\omega)\lesssim m/\omega$. It is worth adding here to the considerations already discussed in \cite{Di_Piazza_2016} that the above conclusion about the transverse formation length is already clear if $\xi_0\lesssim 1$. If $\xi_0\gg 1$, however, one can also observe that: 1) according to the findings in \cite{Meuren_2016}, the regions where the pair is most likely produced are those where the transverse dynamical kinetic momenta of the electron and the positron vanish; 2) since the longitudinal formation length is typically a small fraction $1/\xi_0$ of the laser period, the transverse momentum transfer is again of the order of $m$. Following the above discussion, we can approximately evaluate the external field in Eqs. (\ref{dN_2_BW}) and (\ref{dN_e_BW}) at $\bm{x}_{\perp,+}$ such that the phase $\Delta\Phi_{BW}(\bm{x},\bm{x'})$ becomes
\begin{equation}
\label{Delta_Phi_BW_Appr}
\begin{split}
\Delta\Phi_{BW}(\bm{x},\bm{x}')\approx &\frac{\omega}{\varepsilon\varepsilon'}\frac{T_-}{2}m^2-\frac{T_-}{2\omega}\left(\bm{k}_{\perp}-\frac{\omega}{T_-}\bm{x}_{\perp,-}\right)^2\\
&+\frac{T_-}{2\varepsilon}\left[\bm{p}_{\perp}-\frac{\varepsilon}{T_-}\bm{x}_{\perp,-}-\frac{e}{T_-}\int_T^{T'}d\tilde{T} \bm{\mathcal{A}}_{\perp}(\tilde{\bm{x}})\right]^2\\
&-\frac{\omega}{\varepsilon\varepsilon'}\frac{e^2}{2}\left\{\frac{1}{T_-}\left[\int_T^{T'}d\tilde{T} \bm{\mathcal{A}}_{\perp}(\tilde{\bm{x}})\right]^2+\int_T^{T'}d\tilde{T} \bm{\mathcal{A}}^2_{\perp}(\tilde{\bm{x}})\right\},
\end{split}
\end{equation}
where $\tilde{\bm{x}}=(\tilde{T},\bm{x}_{\perp,+})$. The phase in Eq. (\ref{dN_e_BW}) is the same as this expression of $\Delta\Phi_{BW}(\bm{x},\bm{x}')$ except for the term in the second line which has been integrated out with the transverse momentum $\bm{p}_{\perp}$. Under these approximations, the phases in both Eq. (\ref{dN_2_BW}) and Eq. (\ref{dN_e_BW}) contain at the highest quadratic terms in $\bm{x}_{\perp,-}$ and the resulting integrals in $\bm{x}_{\perp,-}$ are of Gaussian form. With the help of the identities (\ref{G_0}) and (\ref{G_2}), these integrals can be taken analytically and Eqs. (\ref{dN_2_BW}) and (\ref{dN_e_BW}) become
\begin{equation}
\label{dN_3_BW}
\begin{split}
\frac{dN_{BW}}{d\varepsilon d\Omega_p}=&\frac{\rho_{\Sigma,\gamma}}{8\pi^2}\frac{\alpha\varepsilon}{\omega\varepsilon'}\int dT dT'd^2\bm{x}_{\perp}\\
&\times e^{i\frac{\omega}{2\varepsilon\varepsilon'}\left\langle T_-\left\{m^2+\left[\bm{p}_{\perp}-\frac{\varepsilon}{\omega}\bm{k}_{\perp}-\frac{e}{T_-}\int_T^{T'}d\tilde{T} \bm{\mathcal{A}}_{\perp}(\tilde{\bm{x}})\right]^2\right\}-e^2\left\{\frac{1}{T_-}\left[\int_T^{T'}d\tilde{T} \bm{\mathcal{A}}_{\perp}(\tilde{\bm{x}})\right]^2+\int_T^{T'}d\tilde{T} \bm{\mathcal{A}}^2_{\perp}(\tilde{\bm{x}})\right\}\right\rangle}\\
&\times\left\{ m^2\left(\frac{\varepsilon'}{\varepsilon}+\frac{\varepsilon}{\varepsilon'}+4\right)+\frac{\omega^2}{\varepsilon\varepsilon'}\left[\bm{p}_{\perp}-\frac{\varepsilon}{\omega}\bm{k}_{\perp}+e\bm{\mathcal{A}}_{\perp,+}(\bm{x},\bm{x}')\right]^2\right.\\
&\left.-e^2\frac{(\varepsilon-\varepsilon')^2}{4\varepsilon\varepsilon'}\bm{\mathcal{A}}^2_{\perp,-}(\bm{x},\bm{x}')\right\},
\end{split}
\end{equation}
and
\begin{equation}
\label{dN_e_2_BW}
\begin{split}
\frac{dN_{BW}}{d\varepsilon}=&i\frac{\rho_{\Sigma,\gamma}}{4\pi}\frac{\alpha}{\omega^2}\int \frac{dT dT'd^2\bm{x}_{\perp}}{T_-} e^{i\frac{\omega}{2\varepsilon\varepsilon'}\left\langle m^2T_--e^2\left\{\frac{1}{T_-}\left[\int_T^{T'}d\tilde{T} \bm{\mathcal{A}}_{\perp}(\tilde{\bm{x}})\right]^2+\int_T^{T'}d\tilde{T} \bm{\mathcal{A}}^2_{\perp}(\tilde{\bm{x}})\right\}\right\rangle}\\
&\times\Bigg\{m^2\left(\frac{\varepsilon'}{\varepsilon}+\frac{\varepsilon}{\varepsilon'}+4\right)+\frac{2i\omega}{T_-}+\frac{\omega^2}{\varepsilon\varepsilon'}e^2\left[\frac{1}{T_-}\int_T^{T'}d\tilde{T} \bm{\mathcal{A}}_{\perp}(\tilde{\bm{x}})+\bm{\mathcal{A}}_{\perp,+}(\bm{x},\bm{x}')\right]^2\\
&-e^2\frac{(\varepsilon-\varepsilon')^2}{4\varepsilon\varepsilon'}\bm{\mathcal{A}}^2_{\perp,-}(\bm{x},\bm{x}')\Bigg\},
\end{split}
\end{equation}
respectively. Following the discussion below Eq. (\ref{dN_e_BW}), we point out that the symbols $\tilde{\bm{x}}$ and $\bm{x}'$ have to be intended here as $\tilde{\bm{x}}=(\tilde{T},\bm{x}_{\perp})$ and $\bm{x}'=(T',\bm{x}_{\perp})$, respectively. It is worth noticing that neglecting the difference between $\bm{x}_{\perp}$ and $\bm{x}'_{\perp}$ in the external vector potential and evaluating it at the average transverse coordinate $\bm{x}_{\perp,+}$ implies that the resulting expressions (\ref{dN_3_BW}) and (\ref{dN_e_2_BW}) have the same form as in a plane wave with four-vector potential $A^{\mu}_{PW,\perp}(T)=(0,\bm{A}_{PW,\perp}(T),0)$ with the substitution $\bm{A}_{PW,\perp}(T)\to \bm{\mathcal{A}}_{\perp}(\bm{x})$. Thus, under our approximations in which we keep leading-order terms in $1/\omega$, the transverse conjugated momentum $\bm{p}_{\perp}-e\bm{\mathcal{A}}_{\perp}(\bm{x})$ is approximately conserved, which is consistent with the analysis on the classical dynamics presented in \cite{Di_Piazza_2014} (see also the discussion below at the end of the Section).

Both Eq. (\ref{dN_3_BW}) and Eq. (\ref{dN_e_2_BW}) can be substantially simplified either by performing suitable integrations by parts \cite{Boca_2009,Mackenroth_2011} or equivalently by enforcing gauge invariance with respect to the incoming photon \cite{Ilderton_2011}. In Eq. (\ref{dN_3_BW}) one can then use the fact that 
\begin{equation}
0=\int dT_-\frac{\partial}{\partial T_-} e^{i\frac{\omega}{2\varepsilon\varepsilon'}\left\langle T_-\left\{m^2+\left[\bm{p}_{\perp}-\frac{\varepsilon}{\omega}\bm{k}_{\perp}-\frac{e}{T_-}\int_T^{T'}d\tilde{T} \bm{\mathcal{A}}_{\perp}(\tilde{\bm{x}})\right]^2\right\}-e^2\left\{\frac{1}{T_-}\left[\int_T^{T'}d\tilde{T} \bm{\mathcal{A}}_{\perp}(\tilde{\bm{x}})\right]^2+\int_T^{T'}d\tilde{T} \bm{\mathcal{A}}^2_{\perp}(\tilde{\bm{x}})\right\}\right\rangle},
\end{equation}
where $T=T_++T_-/2$ and $T'=T_+-T_-/2$ and obtain
\begin{equation}
\label{dN_4_BW}
\begin{split}
\frac{dN_{BW}}{d\varepsilon d\Omega_p}=&\frac{\rho_{\Sigma,\gamma}}{4\pi^2}\frac{\alpha m^2\varepsilon}{\omega\varepsilon'}\int dT dT'd^2\bm{x}_{\perp} e^{-i\frac{\omega}{2\varepsilon\varepsilon'}\int_T^{T'}d\tilde{T}\left\{m^2+\left[\bm{p}_{\perp}-\frac{\varepsilon}{\omega}\bm{k}_{\perp}+e\bm{\mathcal{A}}_{\perp}(\tilde{\bm{x}})\right]^2\right\}}\\
&\times\left[1-\frac{e^2}{4}\frac{\varepsilon^2+\varepsilon^{\prime 2}}{\varepsilon\varepsilon'}\frac{\bm{\mathcal{A}}^2_{\perp,-}(\bm{x},\bm{x}')}{m^2}\right].
\end{split}
\end{equation}
From the computational point of view, this expression can be rewritten in a more suitable form by noticing that
\begin{equation}
e^2\bm{\mathcal{A}}^2_{\perp,-}(\bm{x},\bm{x}')=[\bm{\pi}_{\perp,p}(\bm{x})-\bm{\pi}_{\perp,p}(\bm{x}')]^2,
\end{equation}
where $\bm{\pi}_{\perp,p}(\bm{x})=\bm{p}_{\perp}-(\varepsilon/\omega)\bm{k}_{\perp}+e\bm{\mathcal{A}}_{\perp}(\bm{x})$. Now, by integrating by parts the resulting terms proportional to $\bm{\pi}^2_{\perp,p}(\bm{x})$ and $\bm{\pi}^2_{\perp,p}(\bm{x}')$, one can easily show that
\begin{equation}
\label{dN_5_BW}
\begin{split}
\frac{dN_{BW}}{d\varepsilon d\Omega_p}=&\frac{\rho_{\Sigma,\gamma}}{8\pi^2}\frac{\alpha m^2\varepsilon}{\omega\varepsilon'}\int d^2\bm{x}_{\perp} \left[\frac{\omega^2}{\varepsilon\varepsilon'}|f_{0,p}(\bm{x}_{\perp})|^2+\frac{\varepsilon^2+\varepsilon^{\prime\,2}}{\varepsilon\varepsilon'}\left\vert\frac{f_{0,p}(\bm{x}_{\perp})}{m}\left(\bm{p}_{\perp}-\frac{\varepsilon}{\omega}\bm{k}_{\perp}\right)+\bm{f}_{1,p}(\bm{x}_{\perp})\right\vert^2\right],
\end{split}
\end{equation}
where
\begin{align}
f_{0,p}(\bm{x}_{\perp})&=\int dT e^{i\frac{\omega}{2\varepsilon\varepsilon'}\int_0^Td\tilde{T}\left[m^2+\bm{\pi}_{\perp,p}^2(\tilde{\bm{x}})\right]},\\
\bm{f}_{1,p}(\bm{x}_{\perp})&=\frac{e}{m}\int dT \bm{\mathcal{A}}_{\perp}(\bm{x})e^{i\frac{\omega}{2\varepsilon\varepsilon'}\int_0^Td\tilde{T}\left[m^2+\bm{\pi}_{\perp,p}^2(\tilde{\bm{x}})\right]}.
\end{align}
Finally, the integral $f_{0,p}(\bm{x}_{\perp})$ can be regularized as indicated above and one obtains the relation
\begin{equation}
\label{Regularization_BW}
\left[m^2+\left(\bm{p}_{\perp}-\frac{\varepsilon}{\omega}\bm{k}_{\perp}\right)^2\right]f_{0,p}(\bm{x}_{\perp})+2m\left(\bm{p}_{\perp}-\frac{\varepsilon}{\omega}\bm{k}_{\perp}\right)\cdot
\bm{f}_{1,p}(\bm{x}_{\perp})+m^2f_{2,p}(\bm{x}_{\perp})=0,
\end{equation}
where
\begin{equation}
\label{f_2_BW}
f_{2,p}(\bm{x}_{\perp})=\frac{e^2}{m^2}\int dT \bm{\mathcal{A}}^2_{\perp}(\bm{x})e^{i\frac{\omega}{2\varepsilon\varepsilon'}\int_0^Td\tilde{T}\left[m^2+\bm{\pi}_{\perp,p}^2(\tilde{\bm{x}})\right]}.
\end{equation}
In this respect, we notice that the choice of the lower integration limit in the phases in the functions $f_{0,p}(\bm{x}_{\perp})$, $\bm{f}_{1,p}(\bm{x}_{\perp})$, and $f_{2,p}(\bm{x}_{\perp})$ is arbitrary and the value $\tilde{T}=0$ has been chosen for convenience.

Now, by analogously integrating by parts the term $2i\omega/T_-^2$ in Eq. (\ref{dN_e_2_BW}) one easily obtains the total energy spectrum in the form
\begin{equation}
\label{dN_e_3_BW}
\begin{split}
\frac{dN_{BW}}{d\varepsilon}=&i\frac{\rho_{\Sigma,\gamma}}{2\pi}\frac{\alpha m^2}{\omega^2}\int \frac{dT dT'd^2\bm{x}_{\perp}}{T_-} e^{i\frac{\omega}{2\varepsilon\varepsilon'}\left\langle m^2T_--e^2\left\{\frac{1}{T_-}\left[\int_T^{T'}d\tilde{T} \bm{\mathcal{A}}_{\perp}(\tilde{\bm{x}})\right]^2+\int_T^{T'}d\tilde{T} \bm{\mathcal{A}}^2_{\perp}(\tilde{\bm{x}})\right\}\right\rangle}\\
&\times\left[1-\frac{e^2}{4}\frac{\varepsilon^2+\varepsilon^{\prime 2}}{\varepsilon\varepsilon'}\frac{\bm{\mathcal{A}}^2_{\perp,-}(\bm{x},\bm{x}')}{m^2}\right],
\end{split}
\end{equation}
where we recall that the factor $T_-$ in the denominator has to be intended as $T_-\to T_-+i0$ as it results from the condition on the imaginary part of the constant $a$ once one applies the results in the integrals in Eqs. (\ref{G_0}) and (\ref{G_2}). Note that the prescription $T_-\to T_-+i0$ ensures that $dN_{BW}/d\varepsilon$ vanishes for vanishing background field. By comparing Eqs. (\ref{dN_3_BW}) and (\ref{dN_e_2_BW}) with Eq. (9) and Eq. (10), respectively, in \cite{Di_Piazza_2016}, we see that the expressions of the angular resolved energy spectra only differ because the transverse momentum of the positron is shifted here by the quantity $-(\varepsilon/\omega)\bm{k}_{\perp}$, whereas the total positron energy spectra coincide. In this respect, we can already conclude that the results in the quasistatic limit ($\xi_0\gg 1$ in the parameter regions where most of the pairs are produced and for typical quantum photon nonlinearity parameter $\kappa_0=2(\omega/m)F_0/F_{cr}$ of the order of unity) are (see Eqs. (11) and (12) in \cite{Di_Piazza_2016})
\begin{equation}
\label{Angular_xi_l_BW}
\frac{dN_{BW}}{d\varepsilon d\Omega_p}=\rho_{\Sigma,\gamma}\frac{\alpha}{\pi^2\sqrt{3}}\frac{\varepsilon^2}{\omega^2}\int  d^3\bm{x}\,g_p(\bm{x})b(\bm{x})\left[1+\frac{\varepsilon^2+\varepsilon^{\prime\,2}}{\varepsilon\varepsilon'}g_p^2(\bm{x})\right]\text{K}_{1/3}\left(\frac{2}{3}b(\bm{x})g_p^3(\bm{x})\right)
\end{equation}
and
\begin{equation}
\label{Spectral_xi_l_BW}
\frac{dN_{BW}}{d\varepsilon}=\rho_{\Sigma,\gamma}\frac{\alpha}{\pi\sqrt{3}}\frac{m^2}{\omega^2}\int  d^3\bm{x}\left[\frac{\varepsilon^2+\varepsilon^{\prime\,2}}{\varepsilon\varepsilon'}\text{K}_{2/3}\left(\frac{2}{3}b(\bm{x})\right)+\int_{\frac{2}{3}b(\bm{x})}^{\infty}dz\,\text{K}_{1/3}(z)\right],
\end{equation}
where we have introduced the functions $g_p(\bm{x})=\sqrt{1+\bm{\pi}^2_{\perp,p}(\bm{x})/m^2}$ and  $b(\bm{x})=(\omega^2/\varepsilon\varepsilon')\kappa^{-1}(\bm{x})$, with $\kappa(\bm{x})=(\omega/m)|\partial\bm{\mathcal{A}}_{\perp}(\bm{x})/\partial T|/F_{cr}$ being the local value of the quantum photon nonlinearity parameter. Concerning the comparison with the results in \cite{Baier_b_1998}, since now $\bm{k}_{\perp}\neq\bm{0}$ it is more transparent to first observe that there the differential spectra with respect to the electron momenta (energy) are reported (see Eqs. (3.28) and (3.30) in \cite{Baier_b_1998}). If we had integrated over the positron degrees of freedom, we would have obtained the same expression with all primed and unprimed energies and momenta (and consequently solid angle) exchanged and with $\bm{\mathcal{A}}_{\perp}(\bm{x})\to-\bm{\mathcal{A}}_{\perp}(\bm{x})$. Then, one can easily see that the results are in agreement by correctly identifying the electron transverse velocity there with the quantity $[\bm{p}'_{\perp}-e\bm{\mathcal{A}}_{\perp}(\bm{x})]/\varepsilon'$ here. Analogously to what we have mentioned in \cite{Di_Piazza_2016}, the corresponding results for a single incoming photon in a plane wave are formally obtained by removing the dependence on the transverse coordinates from the external field and by setting $\rho_{\Sigma,\gamma}\int d^2\bm{x}_{\perp}=1$. On the other hand, it should be noticed that by starting, for example, from the positron angular distribution in a plane wave expressed in terms of the transverse potential $\bm{A}_{PW,\perp}(T)$, the corresponding result in a focused field is not simply obtained via the substitution $\bm{A}_{PW,\perp}(T)\to \bm{A}_{\perp}(\bm{x})$ and then by averaging over the transverse coordinates but the gauge invariant quantity $\bm{\mathcal{A}}_{\perp}(\bm{x})$ has to be constructed first in terms of the electromagnetic field in the non-plane-wave case (see Eq. (\ref{A_out})).

\section{Nonlinear single Compton scattering}

In order to simplify the comparison with the formulas obtained in the previous section, we assume here that the incoming (outgoing) electron has four-momentum and the spin quantum number $p^{\mu}=(\varepsilon,\bm{p})$ and $s$ ($p^{\prime\mu}=(\varepsilon',\bm{p}')$ and $s'$), respectively. Analogously the emitted photon has four-momentum $k^{\mu}=(\omega,\bm{k})$ and (linear) polarization $l$ (polarization four-vector $e^{\mu}_{k,l}$). The leading-order $S$-matrix element of nonlinear single Compton scattering in the Furry picture reads \cite{Furry_1951,Landau_b_4_1982}
\begin{equation}
\label{S_C}
S_{C,fi}=-ie\sqrt{4\pi}\int d^4x\,\bar{\psi}^{(\text{out})}_{p',s'}(x)\frac{\hat{e}_{k,l}}{\sqrt{2\omega}}e^{i(kx)}\psi^{(\text{in})}_{p,s}(x).
\end{equation}
Under the present conditions the in-state $\psi^{(\text{in})}_{p,s}(x)$ can be written in the form \cite{Di_Piazza_2014,Di_Piazza_2015}
\begin{equation}
\psi^{(\text{in})}_{p,s}(x)=e^{iS^{(\text{in})}_{p}(x)}\bigg[1+\frac{e}{4 p_+}\hat{n}\hat{\mathcal{A}}^{(\text{in})}(\bm{x})\bigg] \frac{u_{p,s}}{\sqrt{2\varepsilon}},
\end{equation}
where
\begin{equation}
\begin{split}
S^{(\text{in})}_{p}(x)=&-(p_+\phi+p_-T-\bm{p}_{\perp}\cdot\bm{x}_{\perp})-e\int_{-\infty}^Td\tilde{T} A_-(\tilde{\bm{x}})\\
&-\frac{1}{p_+}\int_{-\infty}^Td\tilde{T} \left[e(p\mathcal{A}^{(\text{in})}(\tilde{\bm{x}}))-
\frac{1}{2}e^2\mathcal{A}^{(\text{in})\,2}(\tilde{\bm{x}})\right],
\end{split}
\end{equation}
where $\mathcal{A}^{(\text{in}),\mu}(\bm{x})=(0,\bm{\mathcal{A}}^{(\text{in})}_{\perp}(\bm{x}),0)$, with
\begin{equation}
\bm{\mathcal{A}}^{(\text{in})}_{\perp}(\bm{x})=\bm{A}_{\perp}(\bm{x})+\bm{\nabla}_{\perp}\int_{-\infty}^Td\tilde{T}A_-(\tilde{\bm{x}})
=-\int_{-\infty}^Td\tilde{T}[\bm{E}_{\perp}(\tilde{\bm{x}})+\bm{z}\times\bm{B}_{\perp}(\tilde{\bm{x}})].
\end{equation}
As a first important result, we would like to show that within the matrix element we can consistently approximate $\bm{\mathcal{A}}^{(\text{out})}_{\perp}(\bm{x})\approx\bm{\mathcal{A}}^{(\text{in})}_{\perp}(\bm{x})\equiv \bm{\mathcal{A}}_{\perp}(\bm{x})$, such that we can remove the upper index in this quantity for notational simplicity. However, it is more transparent from a physical point of view to use the ``(in)'' expression of $\bm{\mathcal{A}}_{\perp}(\bm{x})$ here because the final results will be expressed in terms of the momentum of the incoming electron. In order to prove the above assertion, we observe that $\Delta\bm{\mathcal{A}}_{\perp}(\bm{x})=\bm{\mathcal{A}}^{(\text{in})}_{\perp}(\bm{x})-\bm{\mathcal{A}}^{(\text{out})}_{\perp}(\bm{x})=\bm{\nabla}_{\perp}\int_{-\infty}^{\infty}d\tilde{T}A_-(\tilde{\bm{x}})\equiv \Delta\bm{\mathcal{A}}_{\perp}(\bm{x}_{\perp})$, with the last equality being justified as under our approximations $\Delta\bm{\mathcal{A}}_{\perp}(\bm{x})$ depends only on the two transverse coordinates. Since the difference $\Delta\bm{\mathcal{A}}_{\perp}(\bm{x}_{\perp})$ is the gradient of a scalar function, it is clear that $\bm{\nabla}_{\perp}\times\Delta\bm{\mathcal{A}}_{\perp}(\bm{x}_{\perp})=\bm{0}$. Now, in general, each component $A^{\mu}(x)$ of the four-vector potential field fulfills the wave equation $2\partial^2A^{\mu}/\partial T\partial\phi-\bm{\nabla}^2_{\perp}A^{\mu}=0$ (recall that we work in the Lorentz gauge). Thus, by integrating the corresponding equation for $A_-(x)$ and realistically assuming that $\lim_{T\to\pm\infty}A_-(x)=0$, we obtain that $\bm{\nabla}_{\perp}\cdot\Delta\bm{\mathcal{A}}_{\perp}(\bm{x}_{\perp})=0$. By exploiting the Helmholtz theorem (see, e.g., \cite{Girault_b_1986}), we can conclude that $\Delta\bm{\mathcal{A}}_{\perp}(\bm{x}_{\perp})=\bm{0}$ as the fields have to vanish at infinity. Consequently, the quantity $\int_{-\infty}^{\infty}d\tilde{T}A_-(\tilde{\bm{x}})$ can also be ignored in the phase of the amplitude in Eq. (\ref{S_C}). This conclusion can also be justified physically in the case of a tightly focused laser beam as being related to the fact that realistic propagating beams do not have dc components.

Now, the number $dN_C$ of photons emitted with momenta between $\bm{k}$ and $\bm{k}+d\bm{k}$ is given by
\begin{equation}
dN_C=\mathcal{N}_eV\frac{d^3\bm{k}}{(2\pi)^3} V\frac{d^3\bm{p}'}{(2\pi)^3} \frac{1}{2}\sum_{l,s,s'}|S_{C,fi}|^2,
\end{equation}
where $\mathcal{N}_e$ is the number of incoming electrons and where for the sake of clarity the quantization volume $V=L_xL_yL_z$ has been explicitly indicated (recall that the $S$-matrix element $S_{C,fi}$ contains a factor $1/V^{3/2}$). Now, since the dependence of the background field on the coordinate $\phi$ can be ignored, the corresponding component of the conjugated momentum is conserved and the resulting $\delta$-function reads $\delta(p'_++k_+-p_+)\approx \delta(p'_z+k_z-p_z)$. By squaring this $\delta$-function we obtain $\delta(p'_++k_+-p_+)^2\approx \delta(p'_z+k_z-p_z)^2\approx (2\pi)^{-1}L_z\delta(p'_z+k_z-p_z)$. Moreover, the sum over the spin variables and over the photon polarization leads to the evaluation of the trace:
\begin{equation}
\begin{split}
\mathcal{T}_C=&-\frac{1}{4}\text{Tr}\left\{(\hat{p}'+m)\bigg[1-\frac{e}{4 p'_+}\hat{n}\hat{\mathcal{A}}(\bm{x})\bigg]
\gamma^{\mu}\bigg[1+\frac{e}{4 p_+}\hat{n}\hat{\mathcal{A}}(\bm{x})\bigg](\hat{p}+m)\right.\\
&\left.\times\bigg[1-\frac{e}{4 p_+}\hat{n}\hat{\mathcal{A}}(\bm{x}')\bigg]\gamma_{\mu}\bigg[1+\frac{e}{4 p'_+}\hat{n}\hat{\mathcal{A}}(\bm{x}')\bigg]\right\}.
\end{split}
\end{equation}
The evaluation of $\mathcal{T}_C$ can be carried out with the standard technique as explained, e.g., in \cite{Landau_b_4_1982} and the result is
\begin{equation}
\begin{split}
\mathcal{T}_C=&m^2\left(\frac{\varepsilon'}{\varepsilon}+\frac{\varepsilon}{\varepsilon'}-4\right)+\frac{\varepsilon'}{\varepsilon}\bm{p}^2_{\perp}-2\bm{p}_{\perp}\cdot\bm{p}'_{\perp}+\frac{\varepsilon}{\varepsilon'}\bm{p}^{\prime\, 2}_{\perp}+e\frac{\varepsilon-\varepsilon'}{\varepsilon\varepsilon'}(\varepsilon'\bm{p}_{\perp}-\varepsilon\bm{p}'_{\perp})\cdot[\bm{\mathcal{A}}_{\perp}(\bm{x})+\bm{\mathcal{A}}_{\perp}(\bm{x}')]\\
&-e^2\left[\bm{\mathcal{A}}^2_{\perp}(\bm{x})+\bm{\mathcal{A}}^2_{\perp}(\bm{x}')-\left(\frac{\varepsilon'}{\varepsilon}+\frac{\varepsilon}{\varepsilon'}\right)\bm{\mathcal{A}}_{\perp}(\bm{x})\cdot\bm{\mathcal{A}}_{\perp}(\bm{x}')\right].
\end{split}
\end{equation}
As expected, one can see that this trace can be obtained from the analogous one in nonlinear Breit-Wheeler pair production with the replacement $p^{\mu}\to -p^{\mu}$ and by changing the overall sign. 

By using the above expression of $\mathcal{T}_C$, the quantity $dN_C$ (after performing the integral over the longitudinal momentum of the final electron by exploiting the discussed $\delta$-function) can be written as
\begin{equation}
\label{dN_1_C}
\begin{split}
dN_C=&\rho_{\Sigma,e}\frac{\pi\alpha}{\omega\varepsilon\varepsilon'}\frac{d^2\bm{p}'_{\perp}}{(2\pi)^2}\frac{d\omega}{2\pi}\frac{d^2\bm{k}_{\perp}}{(2\pi)^2}\int d^3\bm{x}d^3\bm{x}'e^{i[\Phi_C(\bm{x})-\Phi_C(\bm{x}')]}\left\{m^2\left(\frac{\varepsilon'}{\varepsilon}+\frac{\varepsilon}{\varepsilon'}-4\right)+\frac{\varepsilon'}{\varepsilon}\bm{p}^2_{\perp}\right.\\
&-2\bm{p}_{\perp}\cdot\bm{p}'_{\perp}+\frac{\varepsilon}{\varepsilon'}\bm{p}^{\prime\, 2}_{\perp}+e\frac{\omega}{\varepsilon\varepsilon'}(\varepsilon'\bm{p}_{\perp}-\varepsilon\bm{p}'_{\perp})\cdot[\bm{\mathcal{A}}_{\perp}(\bm{x})+\bm{\mathcal{A}}_{\perp}(\bm{x}')]\\
&\left.-e^2\left[\bm{\mathcal{A}}^2_{\perp}(\bm{x})+\bm{\mathcal{A}}^2_{\perp}(\bm{x}')-\left(\frac{\varepsilon'}{\varepsilon}+\frac{\varepsilon}{\varepsilon'}\right)\bm{\mathcal{A}}_{\perp}(\bm{x})\cdot\bm{\mathcal{A}}_{\perp}(\bm{x}')\right]\right\},
\end{split}
\end{equation}
where $\rho_{\Sigma,e}=\mathcal{N}_e/L_xL_y$ is the number of incoming electrons per unit surface, where
\begin{equation}
\begin{split}
\Phi_C(\bm{x})=&\left(\frac{m^2+\bm{p}^{\prime 2}_{\perp}}{2\varepsilon'}+\frac{\bm{k}_{\perp}^2}{2\omega}-\frac{m^2+\bm{p}_{\perp}^2}{2\varepsilon}\right)T-(\bm{p}'_{\perp}+\bm{k}_{\perp}-\bm{p}_{\perp})\cdot\bm{x}_{\perp}+e\frac{\bm{p}'_{\perp}}{\varepsilon'}\cdot\int_T^{\infty}d\tilde{T} \bm{\mathcal{A}}_{\perp}(\tilde{\bm{x}})\\
&+e\frac{\bm{p}_{\perp}}{\varepsilon}\cdot\int_{-\infty}^Td\tilde{T} \bm{\mathcal{A}}_{\perp}(\tilde{\bm{x}})-\frac{1}{\varepsilon'}\frac{e^2}{2}\int_T^{\infty}d\tilde{T} \bm{\mathcal{A}}^2_{\perp}(\tilde{\bm{x}})-\frac{1}{\varepsilon}\frac{e^2}{2}\int_{-\infty}^Td\tilde{T} \bm{\mathcal{A}}^2_{\perp}(\tilde{\bm{x}}),
\end{split}
\end{equation}
and where we have exploited the conservation law $\varepsilon=\varepsilon'+\omega$. Notice that, unlike in the pre-exponent, the phase $\Phi_C(\bm{x})-\Phi_C(\bm{x}')$ cannot be simply obtained from the corresponding one in nonlinear Breit-Wheeler pair production with the substitution rules on the photon and the positron four-momentum, but that one also has to take into account the fact that the in- and the out-states become free states at $-\infty$ and $+\infty$, respectively. From the (at the highest) quadratic dependence of $\Phi_C(\bm{x})$ on $\bm{p}'_{\perp}$, it is clear that the resulting integral is Gaussian and it can easily be taken analytically (see Eqs. (\ref{G_0}) and (\ref{G_2})). In this way, the angular resolved photon energy spectrum $dN_C/d\omega d\Omega_{\gamma}$, where $d\Omega_{\gamma}\approx d^2\bm{k}_{\perp}/\omega^2$, reads:
\begin{equation}
\label{dN_2_C}
\begin{split}
\frac{dN_C}{d\omega d\Omega_{\gamma}}=&i\rho_{\Sigma,\gamma}\frac{\alpha\omega}{16\pi^3\varepsilon}\int \frac{d^3\bm{x}d^3\bm{x}'}{T_-}e^{i\Delta\Phi_C(\bm{x},\bm{x}')}\left\langle m^2\left(\frac{\varepsilon'}{\varepsilon}+\frac{\varepsilon}{\varepsilon'}-4\right)+\frac{2i\varepsilon}{T_-}+\frac{\varepsilon'}{\varepsilon}\bigg\{\bm{p}_{\perp}\right.\\
&-\frac{\varepsilon}{T_-}\Delta\bm{x}_{\perp,e}+e\frac{1}{T_-}\bigg[\int_{-\infty}^Td\tilde{T} \bm{\mathcal{A}}_{\perp}(\tilde{\bm{x}})-\int_{-\infty}^{T'}d\tilde{T}' \bm{\mathcal{A}}_{\perp}(\tilde{\bm{x}}')\bigg]\\
&+e\frac{\varepsilon}{\varepsilon'}\frac{1}{T_-}\left[\int_T^{\infty}d\tilde{T} \bm{\mathcal{A}}_{\perp}(\tilde{\bm{x}})-\int_{T'}^{\infty}d\tilde{T}' \bm{\mathcal{A}}_{\perp}(\tilde{\bm{x}}')\right]\\
&+\frac{\omega}{\varepsilon'}e\bm{\mathcal{A}}_{\perp,+}(\bm{x},\bm{x}')\bigg\}^2-e^2\frac{(\varepsilon+\varepsilon')^2}{4\varepsilon\varepsilon'}\bm{\mathcal{A}}^2_{\perp,-}(\bm{x},\bm{x}')\bigg\rangle,
\end{split}
\end{equation}
where
\begin{equation}
\label{Delta_Phi_C}
\begin{split}
\Delta\Phi_C(\bm{x},\bm{x}')=&\frac{\omega}{\varepsilon\varepsilon'}\frac{T_-}{2}m^2+\frac{T_-}{2\omega}\left(\bm{k}_{\perp}-\frac{\omega}{T_-}\bm{x}_{\perp,-}\right)^2-\frac{T_-}{2\varepsilon}\left(\bm{p}_{\perp}-\frac{\varepsilon}{T_-}\Delta\bm{x}_{\perp,e}\right)^2\\
&+\frac{e^2}{2\varepsilon}\left\{\frac{1}{T_-}\left[\int_{-\infty}^Td\tilde{T} \bm{\mathcal{A}}_{\perp}(\tilde{\bm{x}})-\int_{-\infty}^{T'}d\tilde{T}' \bm{\mathcal{A}}_{\perp}(\tilde{\bm{x}}')\right]^2\right.\\
&-\int_{-\infty}^Td\tilde{T} \bm{\mathcal{A}}^2_{\perp}(\tilde{\bm{x}})+\int_{-\infty}^{T'}d\tilde{T}' \bm{\mathcal{A}}^2_{\perp}(\tilde{\bm{x}}')\Bigg\}\\
&-\frac{e^2}{2\varepsilon'}\left\{\frac{1}{T_-}\left[\int_T^{\infty}d\tilde{T} \bm{\mathcal{A}}_{\perp}(\tilde{\bm{x}})-\int_{T'}^{\infty}d\tilde{T}' \bm{\mathcal{A}}_{\perp}(\tilde{\bm{x}}')\right]^2\right.\\
&+\int_T^{\infty}d\tilde{T} \bm{\mathcal{A}}^2_{\perp}(\tilde{\bm{x}})-\int_{T'}^{\infty}d\tilde{T}' \bm{\mathcal{A}}^2_{\perp}(\tilde{\bm{x}}')\Bigg\},
\end{split}
\end{equation}
and where $\Delta\bm{x}_{\perp,e}=\bm{x}_{\perp,-}+(e/\varepsilon)\big[\int_{-\infty}^Td\tilde{T} \bm{\mathcal{A}}_{\perp}(\tilde{\bm{x}})-\int_{-\infty}^{T'}d\tilde{T}' \bm{\mathcal{A}}_{\perp}(\tilde{\bm{x}}')\big]$. Now, Eq. (\ref{dN_2_C}) can be easily integrated with respect to the final transverse photon momentum $d^2\bm{k}_{\perp}\approx \omega^2d\Omega_{\gamma}$ because the integral is Gaussian and the result for the total photon energy spectrum is
\begin{equation}
\label{dN_e_C}
\begin{split}
\frac{dN_C}{d\omega}=&-\rho_{\Sigma,e}\frac{\alpha}{8\pi^2\varepsilon}\int \frac{d^3\bm{x}d^3\bm{x}'}{T^2_-}\exp\left\langle i\left\{\frac{m^2}{2}\frac{\omega}{\varepsilon\varepsilon'}T_--\frac{T_-}{2\varepsilon}\left(\bm{p}_{\perp}-\frac{\varepsilon}{T_-}\Delta\bm{x}_{\perp,e}\right)^2\right.\right.\\
&+\frac{e^2}{2\varepsilon}\left\{\frac{1}{T_-}\left[\int_{-\infty}^Td\tilde{T} \bm{\mathcal{A}}_{\perp}(\tilde{\bm{x}})-\int_{-\infty}^{T'}d\tilde{T}' \bm{\mathcal{A}}_{\perp}(\tilde{\bm{x}}')\right]^2\right.\\
&-\int_{-\infty}^Td\tilde{T} \bm{\mathcal{A}}^2_{\perp}(\tilde{\bm{x}})+\int_{-\infty}^{T'}d\tilde{T}' \bm{\mathcal{A}}^2_{\perp}(\tilde{\bm{x}}')\Bigg\}\\
&-\frac{e^2}{2\varepsilon'}\left\{\frac{1}{T_-}\left[\int_T^{\infty}d\tilde{T} \bm{\mathcal{A}}_{\perp}(\tilde{\bm{x}})-\int_{T'}^{\infty}d\tilde{T}' \bm{\mathcal{A}}_{\perp}(\tilde{\bm{x}}')\right]^2\right.\\
&\left.+\int_T^{\infty}d\tilde{T} \bm{\mathcal{A}}^2_{\perp}(\tilde{\bm{x}})-\int_{T'}^{\infty}d\tilde{T}' \bm{\mathcal{A}}^2_{\perp}(\tilde{\bm{x}}')\Bigg\}\right\rangle\left\langle m^2\left(\frac{\varepsilon'}{\varepsilon}+\frac{\varepsilon}{\varepsilon'}-4\right)+\frac{2i\omega}{T_-}+\frac{\varepsilon'}{\varepsilon}\bigg\{\bm{p}_{\perp}\right.\\
&-\frac{\varepsilon}{T_-}\Delta\bm{x}_{\perp,e}+e\frac{1}{T_-}\bigg[\int_{-\infty}^Td\tilde{T} \bm{\mathcal{A}}_{\perp}(\tilde{\bm{x}})-\int_{-\infty}^{T'}d\tilde{T}' \bm{\mathcal{A}}_{\perp}(\tilde{\bm{x}}')\bigg]\\
&+e\frac{\varepsilon}{\varepsilon'}\frac{1}{T_-}\left[\int_T^{\infty}d\tilde{T} \bm{\mathcal{A}}_{\perp}(\tilde{\bm{x}})-\int_{T'}^{\infty}d\tilde{T}' \bm{\mathcal{A}}_{\perp}(\tilde{\bm{x}}')\right]+\frac{\omega}{\varepsilon'}e\bm{\mathcal{A}}_{\perp,+}(\bm{x},\bm{x}')\bigg\}^2\\
&-e^2\frac{(\varepsilon+\varepsilon')^2}{4\varepsilon\varepsilon'}\bm{\mathcal{A}}^2_{\perp,-}(\bm{x},\bm{x}')\bigg\rangle,
\end{split}
\end{equation}
The above expressions significantly simplify once one exploits that the transverse formation length also in the case of nonlinear single Compton scattering is of the order of the Compton wavelength such that, having in mind realistic applications employing strong either optical or x-ray lasers, we can neglect the difference between $\bm{x}_{\perp}$ and $\bm{x}'_{\perp}$ in the external field. By proceeding in a completely analogous way as in the case of nonlinear Breit-Wheeler pair production, we can approximately evaluate the external field in Eqs. (\ref{dN_2_C})-(\ref{dN_e_C}) at the average transverse coordinate $\bm{x}_{\perp,+}$. Thus, the phase $\Delta\Phi_C(\bm{x},\bm{x}')$ becomes
\begin{equation}
\label{Delta_Phi_C_Appr}
\begin{split}
\Delta\Phi_C(\bm{x},\bm{x}')\approx&\frac{\omega}{\varepsilon\varepsilon'}\frac{T_-}{2}m^2+\frac{T_-}{2\omega}\left(\bm{k}_{\perp}-\frac{\omega}{T_-}\bm{x}_{\perp,-}\right)^2\\
&-\frac{T_-}{2\varepsilon}\left[\bm{p}_{\perp}-\frac{\varepsilon}{T_-}\bm{x}_{\perp,-}+\frac{e}{T_-}\int_T^{T'}d\tilde{T} \bm{\mathcal{A}}_{\perp}(\tilde{\bm{x}})\right]^2\\
&-\frac{\omega}{\varepsilon\varepsilon'}\frac{e^2}{2}\left\{\frac{1}{T_-}\left[\int_T^{T'}d\tilde{T} \bm{\mathcal{A}}_{\perp}(\tilde{\bm{x}})\right]^2+\int_T^{T'}d\tilde{T} \bm{\mathcal{A}}^2_{\perp}(\tilde{\bm{x}})\right\},
\end{split}
\end{equation}
where $\tilde{\bm{x}}=(\tilde{T},\bm{x}_{\perp,+})$. The phase in Eq. (\ref{dN_e_C}) becomes the same as $\Delta\Phi_C(\bm{x},\bm{x}')$ in Eq. (\ref{Delta_Phi_C_Appr}) except for the term proportional to $[\bm{k}_{\perp}-(\omega/T_-)\bm{x}_{\perp,-}]^2$, which is integrated out in Eq. (\ref{dN_e_C}). Thus, after passing from the variables $\bm{x}_{\perp}$ and $\bm{x}'_{\perp}$ to the variables $\bm{x}_{\perp,-}$ and $\bm{x}_{\perp,+}$, the integrals in $\bm{x}_{\perp,-}$ in Eqs. (\ref{dN_2_C}) and (\ref{dN_e_C}) are Gaussian, the identities (\ref{G_0}) and (\ref{G_2}) can be exploited and we obtain
\begin{equation}
\label{dN_3_C}
\begin{split}
\frac{dN_C}{d\omega d\Omega_{\gamma}}=&\frac{\rho_{\Sigma,e}}{8\pi^2}\frac{\alpha\omega}{\varepsilon\varepsilon'}\int dT dT'd^2\bm{x}_{\perp}\\
&\times e^{i\frac{\omega}{2\varepsilon\varepsilon'}\left\langle T_-\left\{m^2+\left[\bm{p}_{\perp}-\frac{\varepsilon}{\omega}\bm{k}_{\perp}+\frac{e}{T_-}\int_T^{T'}d\tilde{T} \bm{\mathcal{A}}_{\perp}(\tilde{\bm{x}})\right]^2\right\}-e^2\left\{\frac{1}{T_-}\left[\int_T^{T'}d\tilde{T} \bm{\mathcal{A}}_{\perp}(\tilde{\bm{x}})\right]^2+\int_T^{T'}d\tilde{T} \bm{\mathcal{A}}^2_{\perp}(\tilde{\bm{x}})\right\}\right\rangle}\\
&\times\left\{m^2\left(\frac{\varepsilon'}{\varepsilon}+\frac{\varepsilon}{\varepsilon'}-4\right)+\frac{\omega^2}{\varepsilon\varepsilon'}\left[\bm{p}_{\perp}-\frac{\varepsilon}{\omega}\bm{k}_{\perp}-e\bm{\mathcal{A}}_{\perp,+}(\bm{x},\bm{x}')\right]^2\right.\\
&\left.-e^2\frac{(\varepsilon+\varepsilon')^2}{4\varepsilon\varepsilon'}\bm{\mathcal{A}}^2_{\perp,-}(\bm{x},\bm{x}')\right\},
\end{split}
\end{equation}
and
\begin{equation}
\label{dN_e_2_C}
\begin{split}
\frac{dN_C}{d\omega}=&i\frac{\rho_{\Sigma,\gamma}}{4\pi}\frac{\alpha}{\varepsilon^2}\int \frac{dT dT'd^2\bm{x}_{\perp}}{T_-} e^{i\frac{\omega}{2\varepsilon\varepsilon'}\left\langle m^2T_--e^2\left\{\frac{1}{T_-}\left[\int_T^{T'}d\tilde{T} \bm{\mathcal{A}}_{\perp}(\tilde{\bm{x}})\right]^2+\int_T^{T'}d\tilde{T} \bm{\mathcal{A}}^2_{\perp}(\tilde{\bm{x}})\right\}\right\rangle}\\
&\times\Bigg\{m^2\left(\frac{\varepsilon'}{\varepsilon}+\frac{\varepsilon}{\varepsilon'}-4\right)+\frac{2i\omega}{T_-}+\frac{\omega^2}{\varepsilon\varepsilon'}e^2\left[\frac{1}{T_-}\int_T^{T'}d\tilde{T} \bm{\mathcal{A}}_{\perp}(\tilde{\bm{x}})+\bm{\mathcal{A}}_{\perp,+}(\bm{x},\bm{x}')\right]^2\\
&-e^2\frac{(\varepsilon+\varepsilon')^2}{4\varepsilon\varepsilon'}\bm{\mathcal{A}}^2_{\perp,-}(\bm{x},\bm{x}')\Bigg\}.
\end{split}
\end{equation}
Analogously as in Eqs. (\ref{dN_3_BW}) and (\ref{dN_e_2_BW}), the symbols $\tilde{\bm{x}}$ and $\bm{x}'$ have to be intended here as $\tilde{\bm{x}}=(\tilde{T},\bm{x}_{\perp})$ and $\bm{x}'=(T',\bm{x}_{\perp})$, respectively.
In this form both the angular resolved energy spectrum $dN_C/d\omega d\Omega_{\gamma}$ and the total energy spectrum $dN_C/d\omega$ can be obtained from the corresponding quantities in nonlinear Breit-Wheeler pair production with the usual substitutions $\varepsilon\to -\varepsilon$, $\bm{p}_{\perp}\to -\bm{p}_{\perp}$, $\omega\to -\omega$, $\bm{k}_{\perp}\to -\bm{k}_{\perp}$, and then by multiplying the whole expression by $-\rho_{\Sigma,e}\omega^2/\rho_{\Sigma,\gamma}\varepsilon^2$. Moreover, these results are in agreement with the corresponding ones in \cite{Baier_b_1998} obtained within the quasiclassical operator approach and expressed via the electron velocity, whose transverse component is given here by $[\bm{p}_{\perp}-e\bm{\mathcal{A}}_{\perp}(\bm{x})]/\varepsilon$. The comparison is easier if one follows the same procedure leading to Eqs. (\ref{dN_4_BW}) and (\ref{dN_e_3_BW}). The resulting equations for nonlinear single Compton scattering clearly read:
\begin{equation}
\label{dN_4_C}
\begin{split}
\frac{dN_C}{d\omega d\Omega_{\gamma}}=&-\frac{\rho_{\Sigma,e}}{4\pi^2}\frac{\alpha m^2\omega}{\varepsilon\varepsilon'}\int dT dT'd^2\bm{x}_{\perp} e^{-i\frac{\omega}{2\varepsilon\varepsilon'}\int_T^{T'}d\tilde{T}\left\{m^2+\left[\bm{p}_{\perp}-\frac{\varepsilon}{\omega}\bm{k}_{\perp}-e\bm{\mathcal{A}}_{\perp}(\tilde{\bm{x}})\right]^2\right\}}\\
&\times\left[1+\frac{e^2}{4}\frac{\varepsilon^2+\varepsilon^{\prime 2}}{\varepsilon\varepsilon'}\frac{\bm{\mathcal{A}}^2_{\perp,-}(\bm{x},\bm{x}')}{m^2}\right]
\end{split}
\end{equation}
and
\begin{equation}
\label{dN_e_3_C}
\begin{split}
\frac{dN_C}{d\omega}=&-i\frac{\rho_{\Sigma,e}}{2\pi}\frac{\alpha m^2}{\varepsilon^2}\int \frac{dT dT'd^2\bm{x}_{\perp}}{T_-} e^{i\frac{\omega}{2\varepsilon\varepsilon'}\left\langle m^2T_--e^2\left\{\frac{1}{T_-}\left[\int_T^{T'}d\tilde{T} \bm{\mathcal{A}}_{\perp}(\tilde{\bm{x}})\right]^2+\int_T^{T'}d\tilde{T} \bm{\mathcal{A}}^2_{\perp}(\tilde{\bm{x}})\right\}\right\rangle}\\
&\times\left[1+\frac{e^2}{4}\frac{\varepsilon^2+\varepsilon^{\prime 2}}{\varepsilon\varepsilon'}\frac{\bm{\mathcal{A}}^2_{\perp,-}(\bm{x},\bm{x}')}{m^2}\right],
\end{split}
\end{equation}
where the factor $T_-$ in the denominator has to be meant as $T_-\to T_-+i0$ (see the discussion below Eq. (\ref{dN_e_3_BW})).

Also in the present case, of course, we can transform Eq. (\ref{dN_4_C}) in a more suitable form for computation, which is analogous to Eq. (\ref{dN_5_BW}):
\begin{equation}
\label{dN_5_C}
\begin{split}
\frac{dN_C}{d\omega d\Omega_{\gamma}}=&\frac{\rho_{\Sigma,e}}{8\pi^2}\frac{\alpha m^2\omega}{\varepsilon\varepsilon'}\int d^2\bm{x}_{\perp} \left[\frac{\omega^2}{\varepsilon\varepsilon'}|f_{0,e}(\bm{x}_{\perp})|^2+\frac{\varepsilon^2+\varepsilon^{\prime\,2}}{\varepsilon\varepsilon'}\left\vert\frac{f_{0,e}(\bm{x}_{\perp})}{m}\left(\bm{p}_{\perp}-\frac{\varepsilon}{\omega}\bm{k}_{\perp}\right)-\bm{f}_{1,e}(\bm{x}_{\perp})\right\vert^2\right],
\end{split}
\end{equation}
where $\bm{\pi}_{\perp,e}(\bm{x})=\bm{p}_{\perp}-(\varepsilon/\omega)\bm{k}_{\perp}-e\bm{\mathcal{A}}_{\perp}(\bm{x})$, where
\begin{align}
f_{0,e}(\bm{x}_{\perp})&=\int dT e^{i\frac{\omega}{2\varepsilon\varepsilon'}\int_0^Td\tilde{T}\left[m^2+\bm{\pi}_{\perp,e}^2(\tilde{\bm{x}})\right]},\\
\bm{f}_{1,e}(\bm{x}_{\perp})&=\frac{e}{m}\int dT \bm{\mathcal{A}}_{\perp}(\bm{x})e^{i\frac{\omega}{2\varepsilon\varepsilon'}\int_0^Td\tilde{T}\left[m^2+\bm{\pi}_{\perp,e}^2(\tilde{\bm{x}})\right]},
\end{align}
and where the quantity $f_{0,e}(\bm{x}_{\perp})$ has to be computed according to the relation
\begin{equation}
\label{Regularization_C}
\left[m^2+\left(\bm{p}_{\perp}-\frac{\varepsilon}{\omega}\bm{k}_{\perp}\right)^2\right]f_{0,e}(\bm{x}_{\perp})-2m\left(\bm{p}_{\perp}-\frac{\varepsilon}{\omega}\bm{k}_{\perp}\right)\cdot
\bm{f}_{1,e}(\bm{x}_{\perp})+m^2f_{2,e}(\bm{x}_{\perp})=0,
\end{equation}
with
\begin{equation}
\label{f_2_C}
f_{2,e}(\bm{x}_{\perp})=\frac{e^2}{m^2}\int dT \bm{\mathcal{A}}^2_{\perp}(\bm{x})e^{i\frac{\omega}{2\varepsilon\varepsilon'}\int_0^Td\tilde{T}\left[m^2+\bm{\pi}_{\perp,e}^2(\tilde{\bm{x}})\right]}.
\end{equation}
Also here the choice of the lower integration limit in the phases in the functions $f_{0,e}(\bm{x}_{\perp})$, $\bm{f}_{1,e}(\bm{x}_{\perp})$, and $f_{2,e}(\bm{x}_{\perp})$ is arbitrary and the value $\tilde{T}=0$ has been chosen for convenience.

Thanks to the above mentioned substitution rules on the positron and the photon four-momentum, we finally report, for the sake of completeness, the corresponding expressions of $dN_C/d\omega d\Omega_{\gamma}$ and of $dN_C/d\omega$ in the quasistatic limit (see Eqs. (\ref{Angular_xi_l_BW}) and (\ref{Spectral_xi_l_BW})):
\begin{equation}
\label{Angular_xi_l_C}
\frac{dN_C}{d\omega d\Omega_{\gamma}}=-\rho_{\Sigma,e}\frac{\alpha}{\pi^2\sqrt{3}}\int  d^3\bm{x}\,g_e(\bm{x})b(\bm{x})\left[1-\frac{\varepsilon^2+\varepsilon^{\prime\,2}}{\varepsilon\varepsilon'}g_e^2(\bm{x})\right]\text{K}_{1/3}\left(\frac{2}{3}b(\bm{x})g_e^3(\bm{x})\right)
\end{equation}
and
\begin{equation}
\label{Spectral_xi_l_C}
\frac{dN_C}{d\omega}=\rho_{\Sigma,e}\frac{\alpha}{\pi\sqrt{3}}\frac{m^2}{\varepsilon^2}\int  d^3\bm{x}\left[\frac{\varepsilon^2+\varepsilon^{\prime\,2}}{\varepsilon\varepsilon'}\text{K}_{2/3}\left(\frac{2}{3}b(\bm{x})\right)-\int_{\frac{2}{3}b(\bm{x})}^{\infty}dz\,\text{K}_{1/3}(z)\right],
\end{equation}
where we recall that $\varepsilon'=\varepsilon-\omega$, where $g_e(\bm{x})=\sqrt{1+\bm{\pi}^2_{\perp,e}(\bm{x})/m^2}$, and where $b(\bm{x})=(\omega/\varepsilon')\chi^{-1}(\bm{x})$, with $\chi(\bm{x})=(\varepsilon/m)|\partial\bm{\mathcal{A}}_{\perp}(\bm{x})/\partial T|/F_{cr}$ being the local value of the quantum electron nonlinearity parameter (note that the expression of $b(\bm{x})$ is the same as for nonlinear Breit-Wheeler pair production and this is why we have used the same symbol).

\section{Conclusions}

In conclusion, we have completed the investigation of the experimentally most relevant single-vertex strong field QED processes in a tightly focused laser beam: nonlinear Breit-Wheeler pair production and nonlinear single Compton scattering. The study of the former process was already started in \cite{Di_Piazza_2016} and we have extended here the results obtained there by including the possibility that the incoming photon is not exactly counterpropagating with respect to the laser beam. Moreover, we have determined compact analytical integral expressions of the angular resolved and the total photon energy spectrum in nonlinear single Compton scattering in a tightly focused laser field. Analogously as in \cite{Di_Piazza_2016}, we have exploited the useful approximation that the energy of the incoming electron is the largest dynamical energy of the problem such that the electron is only barely deflected by the laser field, under the assumption that the incoming electron is almost counterpropagating with respect to the laser field. Finally, we have also elucidated the crossing relation between nonlinear Breit-Wheeler pair production and nonlinear single Compton scattering in the presence of a tightly focused laser field. 

\appendix

\section{A possible choice for the background electromagnetic laser field}
Here, we report here the concrete expression of a possible form of the electromagnetic field of the laser, which can be studied with the present formalism. We recall that we consider a laser beam whose focal plane corresponds to the $x\text{-}y$ plane and whose wave vector at the center of the focal area points along the negative $z$ direction. Also, we refer to the Gaussian beam model of a traveling wave as described in \cite{Salamin_2007} and whose four-vector potential $A^{\mu}(x)$ is a solution of Maxwell's equation in vacuum $\partial_{\mu}\partial^{\mu}A^{\nu}(x)=0$ in the Lorentz gauge $\partial_{\mu}A^{\mu}(x)=0$. By indicating as $\omega_0$ the central angular frequency of the laser and by assuming that the laser is linearly polarized along the $x$ direction, the four-vector potential $A^{\mu}(x)$ can be written as the real part of the complex four-vector potential $A^{\mu}_c(x)=(\Phi_c(x),\bm{A}_c(x))$, written in the form
\begin{align}
\Phi_c(x)&=\varphi_c(\bm{r})g(t+z)e^{i\omega_0(t+z)},\\
\bm{A}_c(x)&=\hat{\bm{x}}A_0\psi_c(\bm{r})g(t+z)e^{i\omega_0(t+z)},
\end{align}
where $\bm{r}$ indicates the space coordinates and $A_0=-E_0/\omega_0$. The scalar potential $\Phi_c(x)$ can be written as $\Phi_c(x)=-(i/\omega_0)\bm{\nabla}\cdot\bm{A}_c(x)$ by exploiting the Lorentz gauge condition, whereas the function $\psi_c(\bm{r})$ is found to fulfill the equation $\bm{\nabla}^2\psi_c(\bm{r})+2i\omega_0\partial\psi_c(\bm{r})/\partial z=0$ by imposing that $A^{\mu}(x)$ fulfills the Maxwell's equations in vacuum (in the Lorentz gauge) $\partial_{\mu}\partial^{\mu}A^{\nu}(x)=0$, and the pulse shape function $g(t+z)$ is assumed to be an arbitrary function but slowly-varying on a laser central period $2\pi/\omega_0$ (see also \cite{Salamin_2002}). A possible convenient choice of the function $g(t+z)$ is given by $g(t+z)=\cos^2(\omega_0(t+z)/2N_L)$ for $\omega_0(t+z)\in [-N_L\pi,N_L\pi]$ and $g(t+z)=0$ elsewhere, where $N_L$ corresponds to the number of laser cycles, which is assumed to be much larger than unity. As it is explained in \cite{Salamin_2007}, having in mind the case of a focused laser beam with Gaussian transverse spatial profile it is convenient to express the function $\psi_c(\bm{r})$ as a series $\psi_c(\bm{r})=\sum_{n=0}^{\infty}\psi_{c,2n}(\bm{r})\epsilon_d^{2n}$ in the diffraction angle $\epsilon_d=w_0/z_R$, where $w_0$ is the laser waist size and $z_R=\omega_0w_0^2/2$ is the Rayleigh length. In fact, even for a tightly focused laser beam $w_0\approx\lambda_0$, it is $\epsilon_d\approx 1/\pi\approx 0.3$. In \cite{Salamin_2007} one can find the expression of $\psi_c(\bm{r})$ up to terms of the order $\epsilon_d^{10}$ and the corresponding electromagnetic fields. Here, it is sufficient to report the terms up to $\epsilon_d^4$ \cite{Salamin_2007}:
\begin{align}
\psi_{c,0}(\bm{r})&=fe^{-f\varrho^2},\\
\psi_{c,2}(\bm{r})&=\frac{1}{2}\left(1-\frac{f^2\varrho^4}{2}\right)f^2e^{-f\varrho^2},\\
\psi_{c,4}(\bm{r})&=\frac{1}{8}\left(3-\frac{3f^2\varrho^4}{2}-f^3\varrho^6+\frac{f^4\varrho^8}{4}\right)f^3e^{-f\varrho^2},
\end{align}
where $f=i/(i-\zeta)$ and where the dimensionless variables $\eta=x/w_0$, $\theta=y/w_0$, and $\zeta=z/z_R$ ($\varrho=\sqrt{\eta^2+\theta^2}$) are employed. Analogously, the electric and magnetic field of the laser can be written as the real parts of the complex fields $\bm{E}_c(\bm{r})g(t+z)\exp[i\omega_0(t+z)]$ and $\bm{B}_c(\bm{r})g(t+z)\exp[i\omega_0(t+z)]$. The components of the latter fields are obtained via the equations $\bm{E}_c(\bm{r})=-i\omega_0\bm{A}_c(\bm{r})+(i/\omega_0)\bm{\nabla}(\bm{\nabla}\cdot\bm{A}_c(\bm{r}))$ and $\bm{B}_c(\bm{r})=\bm{\nabla}\times\bm{A}_c(\bm{r})$ \cite{Salamin_2007}. Their precise expressions are quite cumbersome and we refer to the original reference for them \cite{Salamin_2007}. Here, we summarize the main features of the electric and magnetic field and notice that at the lowest order in $\epsilon_d$ the electric field is directed along the $x$ direction and the magnetic field along the $y$ direction, corresponding to the local plane-wave approximation. Linear corrections in $\epsilon_d$ induce the appearance of longitudinal components of the electric and the magnetic field, whereas the $y$ component of the electric field scales as $\epsilon_d^2$. Finally, the $x$ component of the magnetic field vanishes identically. These considerations together with the expressions above of the functions $\psi_{c,2j}(\bm{r})$, with $j=0,1,2$, and $g(t+z)$ show that the spacetime extension of the laser field is determined by $w_0$ on the transverse $x\text{-}y$ plane, and by $\tau=N_L\tau_0$ in time, with $\tau_0=2\pi/\omega_0$ being the central laser period. In order to have an intuition of the extension of the field along the $z$ direction, we notice that the function $\psi_c(\bm{r})$ decreases only linearly along the $z$ direction (for large values of $|\zeta|$) and one cannot rigorously characterize the longitudinal extension of the laser field via the Rayleigh length $z_R$ (or twice its value considering the symmetry of the space structure of the field when changing $z$ to $-z$). Intuitively, one can imagine the laser field as a pulse of length $\tau$ which goes from $z=\infty$ (at $t=-\infty$) to $z=-\infty$ (at $t=+\infty$) and whose peak increases, reaches its maximum when the pulse reaches the plane $z=0$ (at $t=0$), and then again decreases. In order to have a more specific idea of the spacetime extension of the laser beam, we refer to the experimental relevant case of a tightly focused ($w_0\approx \lambda_0$) Ti:Sapphire ($\lambda_0\approx 0.8\;\text{$\mu$m}$) laser beam, which is customarily employed in high-field applications. The focal area on the transverse plane of such a laser beam is of the order of a few square micrometers whereas pulses of about ten cycles, corresponding to $\approx 30\;\text{fs}$ are usually available experimentally (see, e.g., \cite{Di_Piazza_2012}).
%Linux
%\bibliography{/home/theo/tonywolf/Samba/Travagghiu/Bibliography/Books,/home/theo/tonywolf/Samba/Travagghiu/Bibliography/Reviews,/home/theo/tonywolf/Samba/Travagghiu/Bibliography/Papers_Radiation,/home/theo/tonywolf/Samba/Travagghiu/Bibliography/Papers_RR,/home/theo/tonywolf/Samba/Travagghiu/Bibliography/Papers_PP_and_Cascades,/home/theo/tonywolf/Samba/Travagghiu/Bibliography/Papers_VPE,/home/theo/tonywolf/Samba/Travagghiu/Bibliography/Papers_Crystal,/home/theo/tonywolf/Samba/Travagghiu/Bibliography/Papers_Various}
%Windows
%\bibliography{E:/Travagghiu/Bibliography/Books,E:/Travagghiu/Bibliography/Reviews,E:/Travagghiu/Bibliography/Papers_Radiation,E:/Travagghiu/Bibliography/Papers_RR,E:/Travagghiu/Bibliography/Papers_PP_and_Cascades,E:/Travagghiu/Bibliography/Papers_VPE,E:/Travagghiu/Bibliography/Papers_Crystal,E:/Travagghiu/Bibliography/Papers_Various}


\end{document}